\documentclass[journal]{IEEEtran}

\newcommand{\beq}{\begin{equation}}
\newcommand{\eeq}{\end{equation}}
\newcommand{\bsq}{\begin{subequations}}
	\newcommand{\esq}{\end{subequations}}
\newcommand{\bq}{\begin{eqnarray}}
\newcommand{\eq}{\end{eqnarray}}
\newcommand{\bqn}{\begin{eqnarray*}}
	\newcommand{\eqn}{\end{eqnarray*}}
\usepackage{bbm}
\usepackage[pdftex]{graphicx}
\graphicspath{{Figs/}}
\usepackage{enumerate}
\usepackage{enumitem} 

\usepackage{mathptmx}       % selects Times Roman as basic font
\DeclareMathAlphabet{\mathcal}{OMS}{cmsy}{m}{n}
\usepackage{helvet}         % selects Helvetica as sans-serif font
\usepackage{courier}        % selects Courier as typewriter font
\usepackage{type1cm}        % activate if the above 3 fonts are
\usepackage{makeidx}         % allows index generation
\usepackage{graphicx}        % standard LaTeX graphics tool

\usepackage{multicol}        % used for the two-column index
\usepackage{multirow}
\usepackage[bottom]{footmisc}% places footnotes at page bottom
\usepackage{ifthen}
\usepackage{subfigure}
\usepackage[usenames,dvipsnames]{color}

\makeindex             % used for the subject index
\usepackage{graphics} % for pdf, bitmapped graphics files
\usepackage{graphicx} % for pdf, bitmapped graphics files
\usepackage{epsfig} % for postscript graphics files
\usepackage{epstopdf}
\usepackage{mathptmx} % assumes new font selection scheme installed
\usepackage{times} % assumes new font selection scheme installed
\usepackage[cmex10]{amsmath}

\usepackage{mathtools}  % loads amsmath
\usepackage{amssymb}  % assumes amsmath package installed
\usepackage{amsthm}

\usepackage{amsfonts}
\usepackage{cite}
\usepackage{verbatim}
\usepackage{comment}
\usepackage{algorithmic}
\usepackage{multirow}
\usepackage{cite}
\usepackage{stfloats}
\usepackage{supertabular}
\usepackage{longtable}
\usepackage{amssymb}

\DeclareGraphicsExtensions{.pdf,.png,.jpg,.eps,.ps}
\renewcommand{\arraystretch}{1.2}
\usepackage{arydshln}
\usepackage{multicol}
\usepackage{colortbl}
\theoremstyle{definition}

\theoremstyle{definition}
\newtheorem{definition}{Definition}

\ifCLASSINFOpdf
\else
\fi

\usepackage{comment}
\usepackage{ifthen}
\newboolean{showcomments}
\setboolean{showcomments}{true}
\newcommand{\ychen}[1]{\ifthenelse{\boolean{showcomments}}
	{ \textcolor{red}{YC: #1}}}
\newcommand{\tongxin}[1]{\ifthenelse{\boolean{showcomments}}
	{ \textcolor{blue}{(#1)}}{}}

\usepackage{amsmath}
\allowdisplaybreaks[4]

\usepackage[ruled]{algorithm2e}

\begin{document}
	
	\title{Peer-to-Peer Energy Sharing: A New Business Model Towards a Low-Carbon Future}
	
\author{Yue Chen,
        Changhong Zhao
		% <-this % stops a space
	\thanks{Y. Chen is with the Department of Mechanical and Automation Engineering, C. Zhao is with the Department of Information Engineering, the Chinese University of Hong Kong, Hong Kong SAR. (emails: yuechen@mae.cuhk.edu.hk, chzhao@ie.cuhk.edu.hk)}
	\thanks{This work has been submitted to the IEEE for possible publication. Copyright may be transferred without notice, after which this version may no longer be accessible.}}
	

	\markboth{Journal of \LaTeX\ Class Files,~Vol.~XX, No.~X, Feb.~2019}%
	{Shell \MakeLowercase{\textit{et al.}}: Bare Demo of IEEEtran.cls for IEEE Journals}



	\maketitle
	
	\begin{abstract}
The development of distributed generation technology is endowing consumers the ability to produce energy and transforming them into “prosumers”. This transformation shall improve energy efficiency and pave the way to a low-carbon future. 
However, it also exerts critical challenges on system operations, such as the wasted backups for volatile renewable generation and the difficulty to predict behavior of prosumers with conflicting interests and privacy concerns. 
An emerging business model to tackle these challenges is peer-to-peer energy sharing, whose concepts, structures, applications, models, and designs are thoroughly reviewed in this paper, with an outlook of future research to better realize its potentials.
\end{abstract}
	
	\begin{IEEEkeywords}
energy sharing, peer-to-peer energy trading, business model, mechanism design
	\end{IEEEkeywords}

	% \fi
	\IEEEpeerreviewmaketitle

\section{Introduction}\label{sec:intro}
Global warming and climate change are urging governments, industry, and academia to build a low carbon society \cite{zaelke1989global}. Meanwhile, the development of economy and technology makes energy demand grow rapidly with diversifying usage scenarios \cite{BPreport}. This dilemma of \emph{less carbon} versus \emph{more energy} motivates researchers, engineers, and policy-makers to restructure a more sustainable and affordable energy landscape.

A pivotal piece of the puzzle assembling such a landscape is the proliferation of distributed energy resources (DERs) such as small wind turbines, rooftop solar photovoltaic (PV) panels, and energy storage \cite{parag2016electricity}. 
Over 81,000 distributed wind turbines with more than 1GW total capacity were installed in the U.S. during 2003 to 2007 \cite{orrell20162015}. Global residential PV panels increased from 3.7GW in 2004 to 150GW in 2014 \cite{agnew2015effect}. Worldwide battery storage capacity is expected to grow from 2GW  in 2017 to 235GW in 2030 \cite{ralon2017electricity}. These low-carbon technologies demonstrate great potential in relieving environmental pressure, yet accompanied by nontrivial challenges. 

First, the engagement of DERs still relies heavily on policies at this stage. Indeed, deceleration of DER installation has been observed in regions with declining financial supports. For instance, U.K. cut its subsidy for PV adoption by 80\% from 2017 to 2018 \cite{li2018big}. The global storage market shrank by 70\% in 2017--2018 due to the lack of incentives \cite{IEAreport}. 
%The lack of incentives hinders the acceleration of renewable sources towards a greener grid.

Second, the ubiquitous deployment of volatile renewable generation injects uncertainty to the grid and thus calls for increased backups to maintain system reliability. In Japan, many PV owners purchase batteries to hedge against renewable uncertainties and mitigate financial losses \cite{IEAreport}. However, sufficient backups often imply over-redundancy and waste for normal operations, as extreme events rarely occur. 

Third, prevalent DER technologies endow traditional consumers with the ability to adapt demand and produce energy, turning them into responsive loads or prosumers \cite{parag2016electricity}. This brings about tremendous demand-side flexibility that, if fully utilized, can effectively accommodate uncertain renewable generation. However, entities harnessing such flexibility may hold conflicting interests and refrain from cooperation, not to mention their behavior is hard to predict due to privacy. Idle resources and unmet demands can paradoxically coexist as a result of such unpredictability on both sides. 

In common practice, the excessive renewable energy is recycled by the main grid at low prices due to its unstable output. But consider that the renewable energy sources have nearly zero costs, they can thus offer competitive prices to incentivize demand-side flexibility for backup dismissal and efficiency improvement.
%Indeed, it is also this practice that provides a basis to tackle the same challenges, as renewable energy sources can thus offer competitive prices to incentivize demand-side flexibility for backup dismissal and efficiency improvement. 
To unlock this opportunity, the grid operation shall be migrated from a vertically structured market, which integrates transmission (the main grid) and distribution (the local grid), to a horizontally structured \emph{distribution-layer market}, which co-optimizes renewable energy sources, responsive loads, and other DERs without subsidy and without jeopardizing system reliability. 

What inspires such a market design is the emerging sharing economy in other sectors, which is motivated by similar needs to enhance resource utilization and facilitated by advanced information and communication technologies \cite{filippas2020owning}.  
The widespread deployment of those technologies (e.g., smart meters) also makes sharing economy ready for power and energy systems.
In this context, we provide a comprehensive and in-depth review of \emph{energy sharing}---the first of its kind to our knowledge---on its definition, application scenarios, business models, mechanism designs, and future research directions.

TABLE \ref{tab:comparison-review} compares topics discussed by this paper and previous literature reviews, including those for sharing economy in other sectors \cite{ritter2019sharing,cohen2016making}, peer-to-peer energy trading \cite{zhang2017review,sousa2019peer,tushar2020peer,zhou2020state,tushar2021peer,van2018peer,park2017comparative}, and transactive energy \cite{chen2017demand,abrishambaf2019towards}. These reviews adopt different terminology for energy sharing in a broad sense, partly due to the fact that ``sharing'' has a variety of semantic meanings with different conceptual emphases, such as communicating (sharing information), acting together (sharing responsibility), or giving out (sharing a cake). Such ambiguity is intensified by the intangible feature of energy, which differs from tangible assets such as cars and houses in that one does not need physical access to a generator to use energy. As a result, the distinction between energy trading/transactions and energy sharing is not as clear as that between trading and sharing a car or a house. To the best of our knowledge, this review is the first that systematically summarizes the energy-sharing business models and their counterparts in general sharing economy. Moreover, the applications of sharing are first categorized by functions rather than objects. What is also unprecedented is that we elaborate features, design methods, and pros and cons of a variety of sharing mechanisms.

\begin{table*}[t]
	\renewcommand{\arraystretch}{1.3}
	\centering
	\caption{Comparison of relevant literature reviews}
	\label{tab:comparison-review}
	\begin{tabular}{cccccccc}
		\hline 
		 & & Application &	Business model & Mechanism design & Market structure & Pilot project & Legal context  \\
		\cline{3-8}
		\multicolumn{2}{c}{\textbf{Our review}} &  \checkmark & \checkmark & \checkmark & \checkmark & & \\
		\hline
		\multirow{2}{*}{Sharing economy in other sectors} & \cite{ritter2019sharing} & & \checkmark & & & & \\
		& \cite{cohen2016making} & & \checkmark & & \checkmark & & \\
		\hline
		\multirow{7}{*}{Peer-to-peer energy trading} & \cite{zhang2017review} & &  & & & \checkmark &  \\
		& \cite{sousa2019peer} & & & & \checkmark & \checkmark & \\
		& \cite{tushar2020peer} & \checkmark & & \checkmark & & &\\
		& \cite{zhou2020state} & & & & \checkmark & \checkmark & \\
		& \cite{tushar2021peer} & \checkmark & & & \checkmark & \checkmark & \\
		& \cite{van2018peer} & & & & & & \checkmark\\
		& \cite{park2017comparative} & & & & \checkmark & \\
		\hline
		\multirow{2}{*}{Transactive energy} & \cite{chen2017demand} &  \checkmark &  & & & \checkmark &  \\
	    & \cite{abrishambaf2019towards} & \checkmark &  & & \checkmark & \checkmark &  \\
		\hline
	\end{tabular}
\end{table*}

\begin{figure}[t]
	\centering
	\includegraphics[width=1\columnwidth]{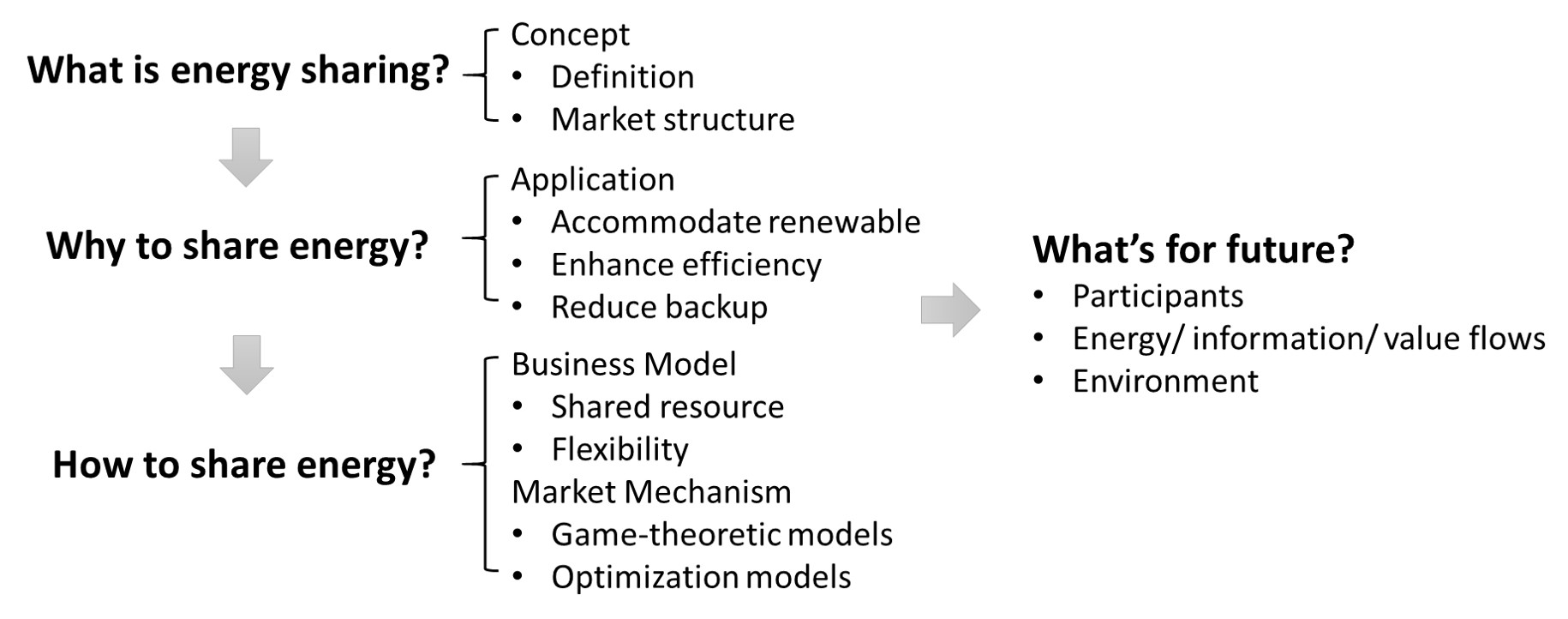}
	\caption{Structure of this literature review.}
	\label{fig:framework}
\end{figure}

Fig. \ref{fig:framework} sketches the structure of this paper. The definition, basic structures, and applications of energy sharing are introduced in Section II; in Section III, business models for energy sharing are categorized by resource sharing modes and flexibility characterizations; 
Section IV traverses energy sharing mechanisms in two categories, i.e., game-theoretic and optimization mechanisms; Section V provides an outlook of future research; and Section VI wraps up the paper. 

\section{Concepts and Applications}
\subsection{Definition and basic structures}
The idea of sharing, initiated from rental markets, is not new, yet its concretization in the energy sector is only boosted after deployment of high-speed information and communication technologies that connect and settle transactions among participants in energy markets. Recent pilot projects have tested energy trading platforms at various geographical scales for different entities, e.g., Piclo \cite{Piclo} and P2P3M \cite{P2P3M} in the U.K., TransActive Grid \cite{mengelkamp2018designing} in the U.S., and Enexa \cite{Enexa} in South Australia. 
A comparison of those pilot projects can be found in \cite{zhou2018evaluation}. Moreover, index systems have been built to evaluate the performance of sharing platforms \cite{zhou2018evaluation,zhou2017performance}.

Essentially, sharing means to utilize otherwise wasted resources or capabilities by transferring their ownership or separating their ownership from the right to use them, targeting a win-win situation across providers and consumers. Inheriting this concept, energy sharing can be defined as follows.

\begin{definition} \textbf{\emph{Energy Sharing}} refers to the business model to optimize energy system operation by acquiring, providing, or sharing access to facilities or energy, leveraging advanced information and communication technologies.
\end{definition}

\begin{figure*}[t]
	\centering
	\includegraphics[width=1.4\columnwidth]{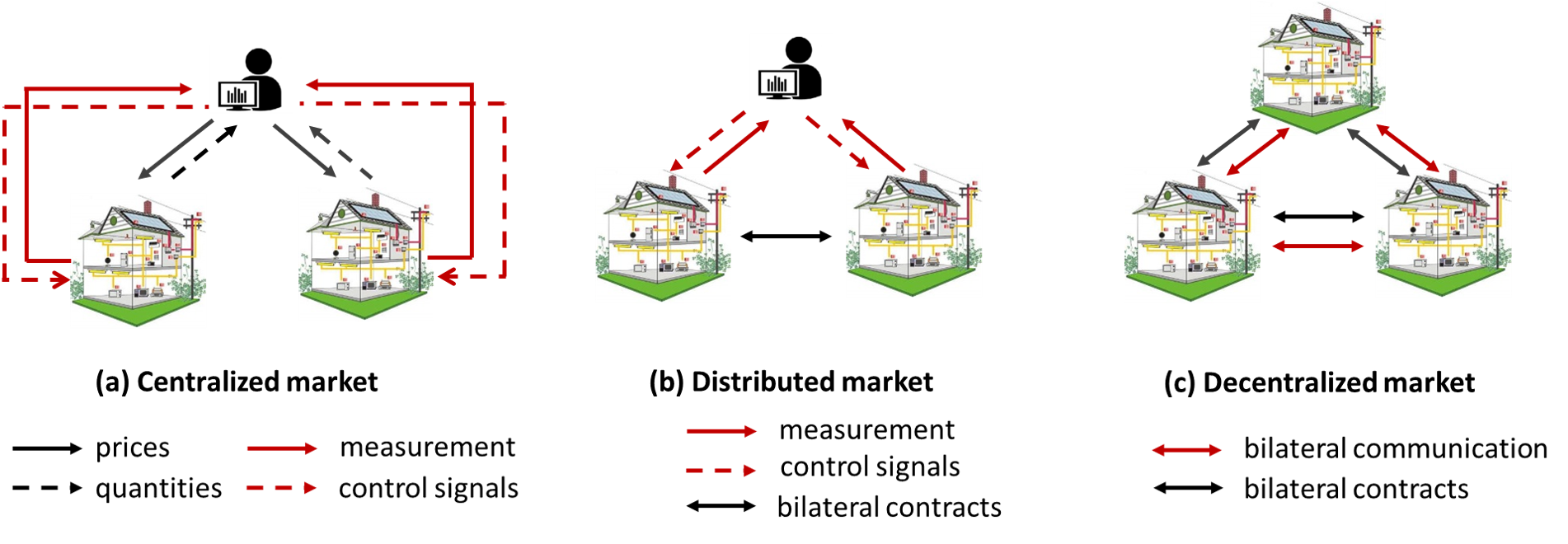}
	\caption{Typical market structures for energy sharing: (a) centralized, (b) distributed, (c) decentralized.}
	\label{fig:market structure}
	\vspace{-0.5em}
\end{figure*}

Market structures for energy sharing generally fall in three categories as shown in Fig. \ref{fig:market structure}.
A centralized market is vertically structured with communications and transactions managed centrally by the operator \cite{chen2018optimal}. Specifically, the operator monitors the status of agents and computes energy prices. Upon receiving the prices, the agents determine and report to the operator their energy quantities to buy or sell. The operator then dispatches control signals to deliver these quantities. 
Most demand-side markets today lie in this category, in which pricing is a big challenge because of privacy concerns and small capacities of agents that make their behavior hard to observe. 
A distributed market settles transactions in a horizontal peer-to-peer (P2P) way. An operator still communicates with all the agents \cite{chen2020approaching}, but instead of directly trading with them, it indirectly influences transactions between them, e.g., by announcing incentive or regulation signals. 
In this way, the agents enjoy a high degree of autonomy in decision making, while preserving their privacy by disclosing limited information to the operator.
In a decentralized market, the agents can communicate and trade between themselves without involving an operator \cite{wang2019distributed}. This structure renders agents the highest levels of flexibility and privacy, but makes it difficult for agents to provide grid services as an aggregate or to maximize their social welfare, due to lack of coordination.

%We would also like to clarify the connections among some frequently used concepts, i.e. local market, peer-to-peer (P2P) market, and energy sharing market. The local market highlights the geographical division, the P2P market focuses on organizational structure, while the energy sharing market puts more emphasis on the purpose. An energy sharing market is usually organized in forms of a local market, but might also be a regional market. The P2P market is a special way to organize an energy sharing market while the vertical market is another way.

\subsection{Applications}
 TABLE \ref{tab:application} categorizes common applications of energy sharing by objectives and scenarios.

\begin{table}[t]
	\renewcommand{\arraystretch}{1.3}
	\centering
	\caption{Summary of energy sharing applications}
	\label{tab:application}
	\begin{tabular}{m{2.5cm}<{\centering}m{2cm}<{\centering}m{2.5cm}<{\centering}}
		\hline 
		Objectives  & Scenarios & References  \\
		\hline
		\multirow{3}{*}{Accommodate renewables} & Solar &  \cite{liu2017energy,liu2018energy,cui2017distributed,fleischhacker2018sharing,long2018aggregated,long2018peer,xu2020two}\\
		&  Wind & \cite{ye2017towards,zhang2018peer,vahedipour2020peer}\\
		& Hydrogen & \cite{tao2020integrated,daneshvar2020transactive,zhu2020energy}\\
		\hline
		\multirow{3}{*}{Enhance efficiency} & Smart building & \cite{alam2019peer,fleischhacker2018sharing,huang2020coordinated,quddus2018collaborative}\\
		&  Microgrid & \cite{chen2020energy,chen2020approaching,ali2020blockchain,akter2020optimal}\\
		& Integrated energy & \cite{yang2020transactive,liu2018hybrid2,zhou2017new,tao2020integrated,zhu2020energy} \\
		\hline
		\multirow{2}{*}{Reduce backups} & Battery & \cite{kalathil2017sharing, lombardi2017sharing,liu2017cloud,he2020community,nguyen2018optimizing,long2018aggregated,long2018peer,rodrigues2020battery}\\
		& Electric vehicle & \cite{he2020utility,sarker2017opportunistic,sun2020blockchain}\\
	    \hline
	\end{tabular}
\end{table}

1) \emph{Accommodate renewable energy}. Renewable generation from solar, wind, hydrogen, etc. can effectively mitigate carbon emissions. However, the intermittent and volatile nature of renewable sources makes it hard for prosumers to plan their energy purchase and usage and hence discourages them from investing in more renewables. 
This difficulty has motivated extensive studies on sharing renewable energy.  
For instance, a group of DERs can mutually cancel out variations in their power outputs to generate a steady aggregate output. In evenings with no sunshine and abundant wind, prosumers with wind turbines can share surplus generation to compensate for the energy shortage of those with PV panels \cite{chen2018analyzing}. 
For solar PV prosumers, prior work developed clustered sharing models \cite{liu2017energy,liu2018energy,xu2020two}, distributed pricing strategies \cite{cui2017distributed}, and resource allocation rules \cite{fleischhacker2018sharing}. 
The performance of solar energy sharing can be further improved by controlling aggregated batteries \cite{long2018aggregated,long2018peer}. With wind power, an online algorithm was proposed to share energy in a cooperative residential community \cite{ye2017towards} and a P2P trading platform was built \cite{zhang2018peer}. Wind energy producers can also trade with demand-response aggregators to co-schedule energy and reserves \cite{vahedipour2020peer}. Moreover, energy sharing models were developed for aggregators with plug-in hybrid electric and hydrogen vehicles \cite{tao2020integrated} and for interconnected microgrids with hydrogen energy storage \cite{daneshvar2020transactive}. The synergy of electricity, hydrogen, and heat networks was studied in \cite{zhu2020energy}.

2) \emph{Enhance operating efficiency}. 
%Energy sharing participants in a building, a microgrid, or a community are equipped with various facilities, such as energy storage and interruptible loads, to make them flexible in trading energy supply and demand. 
Energy sharing can improve the overall efficiency of a group of participants that have different cost or utility functions for their distributed generation, storage, or demand-response units.   
For example, in demand response, a customer can buy energy to compensate for its demand reduction from another customer that incurs a lower cost to reduce the same demand; both customers will gain benefit with a proper payment \cite{chen2020energy}. 
There are three typical scenarios of such efficiency improvement. The first is for smart buildings, e.g., via P2P energy trading among smart homes \cite{alam2019peer} and apartment buildings \cite{fleischhacker2018sharing}, coordinated energy-sharing control in a building cluster \cite{huang2020coordinated}, and energy sharing between commercial buildings, electric vehicle (EV) charging stations, and the grid \cite{quddus2018collaborative}. 
The second scenario is for microgrids, where optimal demand function bidding \cite{chen2020energy,chen2020approaching} and blockchain technology \cite{ali2020blockchain} were used for energy sharing and simulations verified improved efficiency \cite{akter2020optimal}.
The third scenario is for multi-energy systems that integrate power, heat, hydrogen, et cetera \cite{tao2020integrated,zhu2020energy}. 
%are integrated via co-generation facilities such as combined heat and power units, electric boilers, and gas-to-power generators. 
Transactive energy can support economic operations of multi-energy microgrids \cite{yang2020transactive}, building clusters \cite{liu2018hybrid2}, and energy internet \cite{zhou2017new}.

3) \emph{Reduce reliance on backups}. 
%Unpredictability of both supply and demand calls for significant capacities of energy storage as emergency backups, but most of them are never activated before obsolescence as extreme events rarely happen. 
The waste of standby backups raised in Section \ref{sec:intro} can be greatly reduced by energy sharing. %\cite{kontorinis2012managing}. 
Previous research on such backup reduction mainly focused on static batteries and EVs as mobile energy storage. 
The sharing economy brings in new business models for energy storage \cite{kalathil2017sharing, lombardi2017sharing}, among which a representative is cloud storage \cite{liu2017cloud}. 
Indeed, energy storage is commonly co-shared with PVs \cite{nguyen2018optimizing,long2018aggregated,long2018peer}, resting on methods such as adaptive bidding \cite{he2020community}. 
Apart from scheduling, the sizes of batteries were also optimized \cite{rodrigues2020battery}. 
For mobile storage, the potential of energy sharing was revealed by a case study in California \cite{he2020utility}. Game-theoretic approaches were taken to price shared energy between the grid and EVs \cite{sarker2017opportunistic}.  Blockchain technology can also facilitate energy sharing among EVs \cite{sun2020blockchain}.

\section{Business Models}
\subsection{Models by resource sharing modes}

\begin{table*}[t]
	\renewcommand{\arraystretch}{1.3}
	\centering
	\caption{Summary of business models to share resources}
	\label{tab:business_model}
	\begin{tabular}{m{4.5cm}<{\centering}m{5.5cm}<{\centering}m{3cm}<{\centering}m{3cm}<{\centering}}
		\hline 
		Models for sharing resources  & Descriptions & Examples in other sectors & Energy sharing examples \\
		\hline
		Share possessed resources & Operations rely on resources owned by the company or customers.
		% A form of leasing
		%Easy to start. Big investment. 
		& Zipcar, Mobike, Offo & Cloud storage \cite{liu2017cloud,liu2017decision} \\
		Find new homes for used resources & The company acts as an intermediary instead of an owner. & Wallapop, eBay, Peerby & Reusing EV batteries \cite{song2019economy,gur2018reuse,bai2019economic,zhang2020echelon,xu2020study}\\
		Utilize underused resources & Allow owners of idle assets to make money by renting them out. & Rent the Runway, Airbnb, Uber, Blablacar, Shipizy & Sharing solar and wind energy \cite{liu2017energy,liu2018energy,cui2017distributed,fleischhacker2018sharing,long2018aggregated,ye2017towards} \\
		Exploit resource abilities & The more qualified shall fulfill the task to earn with their skills or knowledge. &  TaskRabbit, Zaarly & Demand response \cite{chen2020energy,chen2020approaching}\\
		\hline
	\end{tabular}
\end{table*}
TABLE \ref{tab:business_model} summarizes four models by their modes of resource sharing, with examples from energy and other sectors.

1) \emph{Share possessed resources}. Zipcar, Mobike, and Offo are good examples, which own the vehicles/bikes and track them through web/mobile platforms so that the sharing is essentially leasing \cite{choi2020sustainable}.  
%Specially, a private web-based company owns a cluster of cars (bikes), and each customer registers through an application (APP) to find the closest car (bike), drive (ride) it, and pay for it. The sharing company monitors and tracks all the shared cars (bikes) in real time. 
A drawback of this model is the tremendous financial burden to acquire and maintain the resources to share. An alternative option is to allow each customer to buy such a resource and the company to coordinate them, but this model is hard to initiate due to the scale effect and cross-group externalities \cite{russo2016defining}.
This drawback can be neutralized in energy sharing due to the convenience to access energy across a network, without requiring customer possession of resources.
For example, customers can first to invest in cloud storage, a shared pool of storage centrally controlled by the operator, to receive storage service when needed \cite{liu2017cloud,liu2017decision}.

2) \emph{Find new homes for used resources}. Conventional means such as garage sales to trade used items are restricted by venues and occurrences. 
%eBay was one of the first online marketplaces for global participants to exchange used goods, followed by similar P2P platforms such as Wallapop and Peerby. 
To facilitate cascaded utilization of goods, companies like eBay serve as intermediaries between owners and buyers instead of owning the inventory themselves. 
This model is implemented in energy systems, e.g., with batteries as the goods \cite{song2019economy}. The performance of reused automotive batteries as stationary energy storage has been validated in Europe \cite{gur2018reuse} and China \cite{bai2019economic,zhang2020echelon}. The common practice of replacing lithium-ion EV batteries with lower than 80\% rated capacity is causing serious pollution, while allowing those batteries to participate in energy sharing markets can push forward the sustainable electrification of transportation \cite{xu2020study}. 

3) \emph{Utilize underused resources}. Successful sharing platforms Airbnb and Uber belong to this category. 
For instance, Airbnb brings together owners of vacant rooms and enables them to get cash by posting and renting rooms online. 
%this is just like what Uber does for car owners and Rent the Runway for clothes owners. 
A key difference between the category above (e.g., eBay) and this category is that the former transfers ownership while the latter only temporarily leases the right of use.
In energy systems, intermittency and uncertainty of renewable energy sources force them to be regularly underused to avoid supply-demand mismatch. Energy sharing can improve the utilization of renewable sources by smoothing out their uncertainties, e.g., for PVs \cite{liu2017energy,liu2018energy,cui2017distributed} and other resources \cite{ye2017towards}. Note that this kind of sharing is often aided by batteries \cite{fleischhacker2018sharing,long2018aggregated}.

4) \emph{Exploit resource abilities}. Companies like TaskRabbit assist customers with routine work, e.g., home repair, to find someone better qualified to accomplish the work at a lower cost, so that both the customer and the worker can benefit with a proper payment. 
Similarly, in demand response programs, energy consumption of certain customers can be adjusted at lower disutilities than others, and hence they can perform excessive adjustments to share with those suffering higher disutilities and to reduce the social disutility \cite{chen2020energy,chen2020approaching}.
%Use it or waste it. To buy or to sell, it depends.

\subsection{Models by flexibility characterizations}
Energy sharing models can also be categorized by their flexibility characterizations as explained below. 
Consider agents $i \in \mathcal{I}:=\{1,...,I\}$. Let $q_i$ denote the sharing quantity of agent $i$ and let $\theta_i$ denote its type. An agent $i$ with $q_i>0$ ($q_i<0$) buys (sells) energy from the market. The individual value of agent $i$ is $v_i(q,\theta_i)$, which may depend on all quantities $q:=(q_i, ~\forall i\in \mathcal{I})$. A general formulation of energy sharing is:
\bsq\label{eq:sharing}
\begin{align}
\mathop{\max}_{q_i,\forall i \in \mathcal{I}} ~ & \sum \nolimits_{i \in \mathcal{I}} v_i(q,\theta_i) \label{eq:sharing.1}\\
\mbox{s.t.}~ & \sum \nolimits_{i \in \mathcal{I}} q_i =0 \label{eq:sharing.2}\\
~ & q_i \in \mathcal{D}_i \cap \mathcal{Q}_i,~\forall i \in \mathcal{I}; \quad q \in \tilde{\mathcal{D}} \label{eq:sharing.3}
\end{align}
\esq
which aims to maximize the total value of all the agents. The sharing market is cleared when the total energy sold equals the total bought, i.e.,  \eqref{eq:sharing.2}. The shared energy $q_i$ is limited by individual constraint $\mathcal{D}_i$, such as capacity limit of local resources; network constraint $\tilde{\mathcal{D}}$ that couples the agents; and energy exchange limit $\mathcal{Q}_i$ depending on the sharing model adopted.  The following discussion focuses on $\mathcal{Q}_i$ that characterizes flexibility of the corresponding model.

%In the following, we summarize three mainstream forms of energy sharing with different flexibility levels.
% business-to-crowd, peer-to-peer

1) \emph{Fixed roles, sharing mediating assets}. In this model, shared energy is mediated by certain assets between agents whose roles as sellers or buyers are determined exogenously and beforehand. For example, the sellers charge the cloud storage as a mediating asset, which is then discharged by the buyers \cite{liu2017cloud, liu2017decision}. Flexibility set $\mathcal{Q}_i$ in this case becomes:
\begin{eqnarray}
    0 \le q_i \le Q_i,~\forall i \in \mathcal{B}; \qquad
    -Q_i \le q_i \le 0,~\forall i \in \mathcal{S} \nonumber
\end{eqnarray}
where $\mathcal{B}$ and $\mathcal{S}$ are the set of buyers and sellers, respectively, with $\mathcal{I}=\mathcal{B} \cup \mathcal{S}$. 
%The sets $\mathcal{S}$ and $\mathcal{B}$ are predetermined before the agents enter the market, i.e., their market roles are exogenously given.
Constant $Q_i$ aggregates the physical limits of \emph{mediating assets accessible by agent $i$}, e.g., accessible capacity and maximal charging or discharging rate of cloud storage.

% other examples:Zipcar. 

2) \emph{Fixed roles, sharing local capacity}. Energy may also be traded directly without a mediating asset. For instance, a prosumer with excessive renewable generation in real time may offer it, within local capacity of renewable sources, to one with unexpected demand \cite{liu2017energy,liu2018energy,cui2017distributed}. In this case, $\mathcal{Q}_i$ is:
\begin{eqnarray}
   q_i \ge 0,~\forall i \in \mathcal{B}; \qquad
   q_i \le 0,~\forall i \in \mathcal{S}. \nonumber
\end{eqnarray}
Flexibility of this model is restricted by local capacity $\mathcal{D}_i$ but not any mediating asset. However, the agents are still pre-assigned into a seller set $\mathcal{S}$ or a buyer set $\mathcal{B}$. 

% examples: P2P car-sharing and carpooling models like Blablacar. crowd-shipping models such as Shipizy also belong here.

3) \emph{Flexible roles}. The two models above both pre-assign market roles and hence limit the flexibility of agents, partly due to the rigid feature of shared products. This is similar to the Airbnb model where sellers and buyers are divided by locations of houses. %the specific route in Uber.
This spatial restriction is not stringent for energy that can flow in a network. 
This allows market roles of energy prosumers to be endogenously determined, in which case $\mathcal{Q}_i = \mathbb{R}$ for all $i \in \mathcal{I}$ does not impose additional limits. 
%\begin{align}
%\label{eq:business-model-3}
%    \mathcal{Q}_i = \mathbb{R},~\forall i \in \mathcal{I}.
%\end{align}
References \cite{chen2020energy,chen2020approaching} designed this type of energy sharing mechanisms based on generalized demand function bidding.

As we can observe, the constraint sets $\mathcal{Q}_i$ are expanded and the flexibility is improved as we go through the three models above. Exploiting the flowability of energy, the last model demonstrates the best potential in utilizing flexible resources.
%A solution of \eqref{eq:business-model-1} is always feasible for \eqref{eq:business-model-2}; and a solution of \eqref{eq:business-model-2} works for \eqref{eq:business-model-3}. Therefore, the flexibility of the sharing market improves. 

\section{Mechanism Designs}
Despite the extensive research on general sharing economy \cite{cheng2016sharing}, it is only recently that energy sharing research has grown, started with empirical studies or abstract models to demonstrate its advantages \cite{chen2018analyzing} and followed by mechanism designs that will be reviewed in this section.
%Two key issues for mechanism design are allocation and pricing. 
A good mechanism addresses both pricing and allocation issues to satisfy: 1) The market is cleared with energy supply and demand balanced and network constraints respected. 2) The profit is properly distributed to incentivize customer participation. 3) The mechanism can be implemented with simple rules, without requiring participants to disclose private information.

Let $q=(q_i, ~\forall i\in \mathcal{I})$ be the vector of sharing quantities and $\mathcal{A}_i,~\forall i \in \mathcal{I}$ the action sets of agents; let $\theta_i$ be the type of agent $i$, which is private information from a set $\Theta_i$. Each agent's total value $c_i(.)$ consists of two parts: individual value $v_i(q,\theta_i)$ and monetary transfer $m_i$, where the latter satisfies $\sum \nolimits_{i =1}^I m_i \le 0$. Let $\mathcal{Y}=\{q,m_1,m_2,...,m_I\}$ denote the market outcome. Then a mechanism can be defined as follows. 

\begin{definition} (Mechanism) Let $g: \mathcal{A}_1 \times \mathcal{A}_2 \times ... \times \mathcal{A}_I \to \mathcal{Y}$ be an outcome function. Then a \emph{mechanism} $\Gamma$ is a collection of $\mathcal{A}_i,~\forall i\in \mathcal{I}$ and $g(.)$. A pure strategy for agent $i \in \mathcal{I}$ in mechanism $\Gamma$ is a function $S_i: \Theta_i \to \mathcal{A}_i$ that maps agent types into actions, under which agent $i$ receives value $c_i(g(S),\theta)$.
\end{definition}

\begin{figure}[h]
	\centering
	\includegraphics[width=0.7\columnwidth]{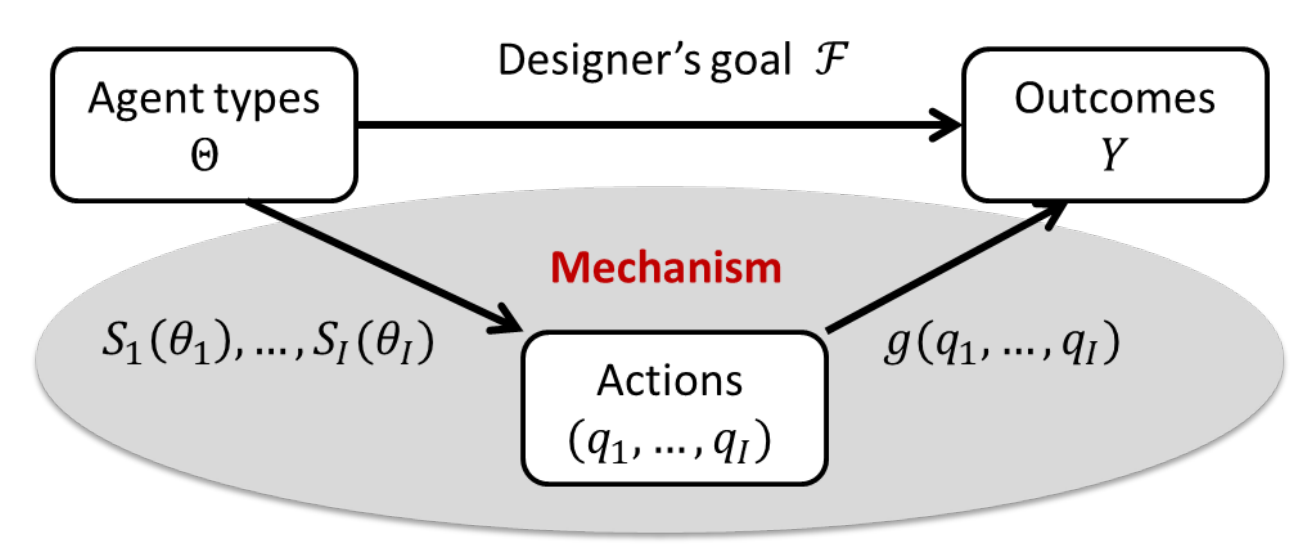}
	\caption{Relationships between designer's goal, mechanism, and outcome.}
	\label{fig:mechanismdef}
\end{figure}

Generally, the mechanism designer has a certain goal $\mathcal{F}:\Theta \to \mathcal{Y}$, e.g., social value maximization. However, since $\theta_i \in \Theta_i$ is private to agent $i \in \mathcal{I}$, the designer can only achieve the goal indirectly via a well-designed mechanism. The relationships between designer's goal, mechanism, and market outcome are shown in Fig. \ref{fig:mechanismdef}.

Existing energy sharing mechanisms can be divided into those based on \emph{game-theoretic models} and \emph{optimization models}. The former category includes cooperative and noncooperative games.  %In the former type, the allocation rule (contracts or operator-set prices) is given first and then each participant reacts to it. The participants are price-takers. In the latter type, the participants are price-makers, and they offer their bids first, and then the sharing market is cleared with prices or matching. 
Cooperative games rest on proper allocation of values, e.g., by Shapley value or nucleolus, to incentivize prosumers to form one or more \emph{coalitions}, which align prosumer interests with social objectives such as minimizing the operational cost of a network or 
maximizing the total utility of prosumers. 
In contrast, noncooperative games focus on competition between individual prosumers. Commonly exploited noncooperative games for energy sharing are Stackelberg games in centralized markets, generalized Nash games in distributed markets, and multi-leader multi-follower games and Nash games in decentralized markets. The latter category includes those using distributed optimization methods and learning algorithms.
TABLE \ref{tab:mechanism} compares these models.
\begin{table*}[t]
	\renewcommand{\arraystretch}{1.3}
	\centering
	\caption{Summary of energy sharing mechanisms}
	\label{tab:mechanism}
	\begin{tabular}{m{1.15cm}<{\centering} | m{2.45cm}<{\centering} | m{2.25cm}<{\centering}m{2.3cm}<{\centering}m{2.3cm}<{\centering}m{2.3cm}<{\centering} | m{2.4cm}<{\centering}}
		\hline 
	 & \multicolumn{5}{c|}{Game-theoretic model} &  \multirow{3}{*}{Optimization model} \\ 
	 \cline{2-6}
		& \multirow{2}{*}{Cooperative game} &  \multicolumn{4}{c|}{Noncooperative game} &  \\
		\cline{3-6}
		&  & Stackelberg game & Generalized Nash game & Multi-leader multi-follower & Bilateral Nash game & \\
		\hline
        Features & Need an allocation rule; agents act cooperatively considering their allocated values. & The operator makes prices, followed by all the agents who decide sharing quantities. & Agents make bids, and the operator returns prices and sharing quantities. & Sellers make prices first, followed by buyers who decide sharing quantities. & Agents act separately as sellers or buyers; sharing quantities are first set and then prices. &  Agents are coordinated by distributed optimization or learning methods. \\
        Structures & Distributed & Centralized & Distributed & Decentralized & Decentralized & Distributed / Decentralized \\
        \hline
        Advantages & Can achieve social optimum without a central coordinator; agents can freely choose to sell or buy. & Can achieve social optimum while satisfying network constraints; budget is balanced. & Agents freely choose to sell or buy; budget balanced; individual interests fulfilled. & No central coordinator needed; budget balanced; individual interests fulfilled.& Can run in an asynchronous manner; budget balanced; individual interests fulfilled. & Can achieve social optimum with network constraints satisfied; may work without coordinator.\\
        Drawbacks  & Hard to maintain budget balance, design allocation rules, and protect privacy. & Need a central coordinator; hard to design prices; agents are price-takers. & Need a coordinator; hard to impose network constraint; market power. & Sellers and buyers are predetermined and hard to match; hard to impose network constraints. & Sellers and buyers are predetermined and hard to match; hard to impose network constraints. & Hard to reveal the economic intuition and incentivize the agents to join.\\
        \hline
        References & Shapley value \cite{mei2019coalitional,han2019estimation,long2019game,liu2018hybrid2}, nucleolus \cite{han2018incentivizing,han2019improving}, Nash bargaining \cite{wang2018incentive,wanghuang2018incentivizing,cui2020community,dutta2014game}, others \cite{chakraborty2018sharing,qi2017sharing,leong2019auction}  &   Applications \cite{liu2017energy,liu2018energy, liu2017energy2, liu2018hybrid, chen2020peer, cui2018two}, proofs \cite{cui2017distributed,tushar2020grid}, algorithms \cite{liu2017online, liu2018distributed,xu2020data}
        & \cite{le2020peer, chen2020energy, chen2020approaching, chen2021energy} & \cite{paudel2018peer, anoh2019energy, lee2015distributed} & \cite{morstyn2019bilateral, oh2020peer, khorasany2020new, chen2018indirect, agate2020enabling, he2020community,wang2014game} & distributed optimization \cite{nguyen2020electric, nguyen2020cooperative, le2018enabling, yang2020transactive, paudel2019pricing}, learning algorithms \cite{chen2019realistic, prasad2019multi}\\
		\hline
	\end{tabular}
\end{table*}

\subsection{Game-theoretic models}

\noindent \textbf{1) Cooperative game-based energy sharing}

\begin{figure}[h]
	\centering
	\includegraphics[width=0.6\columnwidth]{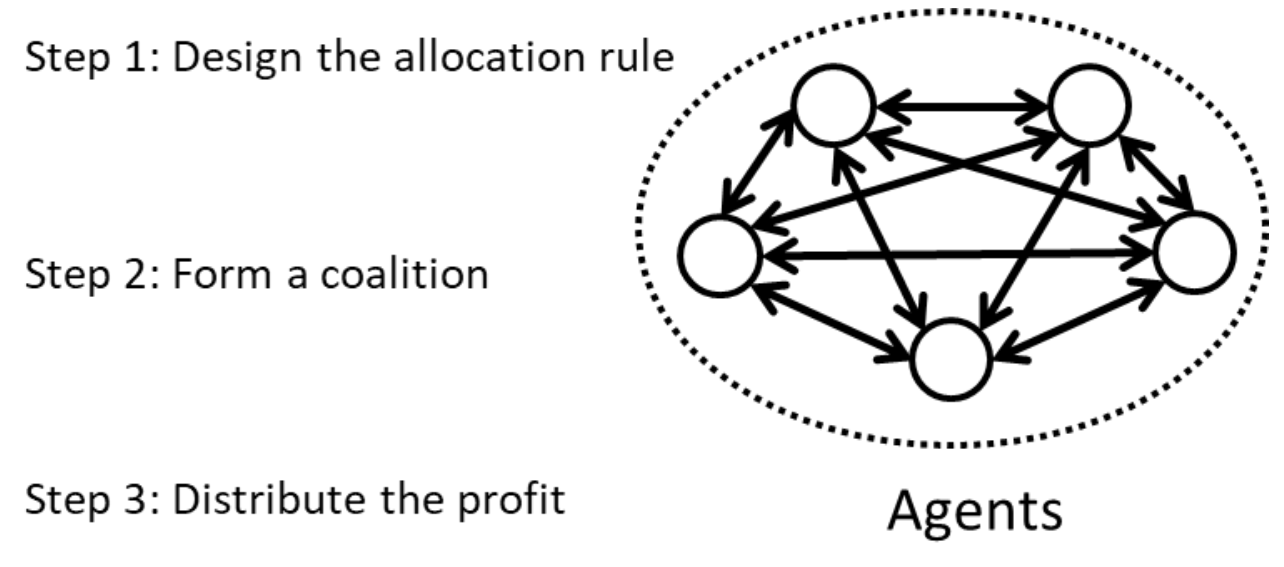}
	\caption{Cooperative game-based energy sharing.}
	\label{fig:category1}
\end{figure}

As Fig. \ref{fig:category1} shows, an allocation rule is designed based on cooperative game theory to achieve certain social objectives, %\cite{branzei2008models}. 
and is implemented by the agents who form coalitions to optimize the value they will each receive. 
Specifically, let $\mathcal{J} \subseteq \mathcal{I}$ denote a coalition (subset) of agents. A game conducted in coalitions can be modeled with a characteristic function $V: 2^I \to \mathbb{R}$ that maps coalitions to values. If the game satisfies superadditivity defined below, then all the agents will cooperate to reach the maximal total value $ V(\mathcal{I})= \sum \nolimits_{i=1}^I v_i(q,\theta_i)$.

\begin{definition} (Superadditive Game) A game with characteristic function $V$ is \emph{superadditive} if $V(\mathcal{J}_1 \cup \mathcal{J}_2) \ge V(\mathcal{J}_1) + V(\mathcal{J}_2)$
for all $\mathcal{J}_1 \subsetneqq \mathcal{I}$, $\mathcal{J}_2 \subsetneqq \mathcal{I}$, $\mathcal{J}_1 \cap \mathcal{J}_2 = \emptyset$.
\end{definition}

Let $c:=(c_i,~\forall i \in \mathcal{I})$ be a value allocation when all the agents collaborate as a \emph{grand coalition}, from which no agent has the incentive to leave if $\sum_{i \in \mathcal{J}} c_i \ge V(\mathcal{J})$ for all $\mathcal{J} \subsetneqq \mathcal{I}$. In this case, we say allocation $c$ is in the \emph{core} of the game. 

%Two well-known approaches to designing a cooperative game are \emph{Shapley value} and \emph{Nucleolus}.
There are two common approaches to the design of cooperative games. First, \emph{Shapley value} distributes values to players according to their marginal contribution, i.e., for player $i$:
\begin{align}\nonumber
    c_i= \sum \limits_{\mathcal{J} \subsetneqq \mathcal{I}/\{i\}} \frac{|\mathcal{J}|!(I-|\mathcal{J}|-1)!}{I!} \left(V(\mathcal{J} \cup \{i\}) - V(\mathcal{J})\right).
\end{align}
With Shapley value, incentives were designed for coalitional operation of networked energy-exchanging microgrids \cite{mei2019coalitional}. Reference \cite{han2019estimation} proposed a random sampling method to estimate the Shapley value of a P2P energy sharing game. Reference \cite{long2019game} compared a Shapley value-based P2P energy trading mechanism with the classical bill sharing, mid-market rate, and supply-demand ratio algorithms. 
A two-level reward allocation scheme was developed to deal with the computational complexity of Shapley value under a large number of agents  \cite{liu2018hybrid2}. Note that Shapley value, albeit a fair allocation, is not always in the core of a cooperative game.

Differently, the \emph{nucleolus} is the set of element(s) in the core that minimizes the overall dissatisfaction of players in the grand coalition measured by the excessive values they would receive should they form alternative (smaller) coalitions. 
Specifically, let $\le_{lex}$ be the \emph{lexicographical} ordering of $\mathbb{R}^n$, i.e., for any $x,y \in \mathbb{R}^n$, there is $x \le_{lex} y$ if either $x=y$ or $\exists ~t \in [0,n)$ such that $x_i=y_i,~\forall i \in [0,t]$ and $x_i < y_i,~\forall i \in [t+1,n]$. 
Denote by $\mathcal{C}$ the set of all possible allocations in the grand coalition, and then the \emph{nucleolus} is the set $\{c_1\in \mathcal{C} ~|~ e(c_1) \le_{lex} e(c_2),\forall c_2 \in \mathcal{C} \}$,
where $e(c)$ is the sequence of excesses $e(\mathcal{J},c)=V(\mathcal{J})-\sum \nolimits_{i \in \mathcal{J}}c_i$ over all the $2^I$ coalitions $\mathcal{J}\subseteq \mathcal{I}$. With the nucleolus, distributed energy storage units can form cost-minimizing coalitions \cite{han2018incentivizing}. A K-means clustering method was used to accelerate the nucleolus calculation \cite{han2019improving}.

To circumvent the possibly heavy computation of Shapley value and nucleolus, \cite{wang2018incentive,wanghuang2018incentivizing,cui2020community,dutta2014game} adopted the \emph{Nash bargaining model}. Ad hoc designs were also made for the sharing of energy storage \cite{chakraborty2018sharing} or among aggregator-managed agents \cite{qi2017sharing}. 
 %effective  cost  reallocation  functions  were  developed  for  storage sharing in smart grid under two scenarios \cite{chakraborty2018sharing}. The cost difference between cooperative sharing and independent operations of agents was split among the aggregator and agents \cite{qi2017sharing}.

The designs so far stick to \emph{budget balance}, i.e., the sum of allocations equals the value of the coalition. An alternative to relax this balance is the \emph{Vickrey-Clarke-Groves (VCG) mechanism}, which pays  $m_i =\sum_{j \ne i} v_j(q(\theta),\theta_j)+h_i(v_{-i})$
to agent $i$, where 
%the first term is the total value of the other agents and
the second term is a function of the (declared) values of agents other than $i$. 
As a result, every agent $i$ chooses its action to maximize $v_i(q,\theta_i)+\sum \nolimits_{j \ne i} v_j(q,\theta_j)$, namely the social value, as $h_i(v_{-i})$ does not depend on agent $i$'s action. The VCG mechanism is proved to be a truthful mechanism in that every agent will report its true value. It was used in \cite{leong2019auction} to limit power losses in P2P energy sharing.

%Although a cooperative game-based energy sharing mechanism can achieve socially optimal efficiency, 
Besides the advantage of achieving social optimality and the concern with computation burden, a serious drawback of cooperative games is that private information, such as value functions of the participants, needs to be disclosed. 
%Moreover, for a large number of participants, calculating the allocation can be time-consuming.

\noindent \textbf{2) Noncooperative game-based energy sharing}

Four noncooperative game models are reviewed below.

\textbf{2.1) Stackelberg game}

\begin{figure}
\centering
\subfigure[]
{\includegraphics[width=0.48\columnwidth]{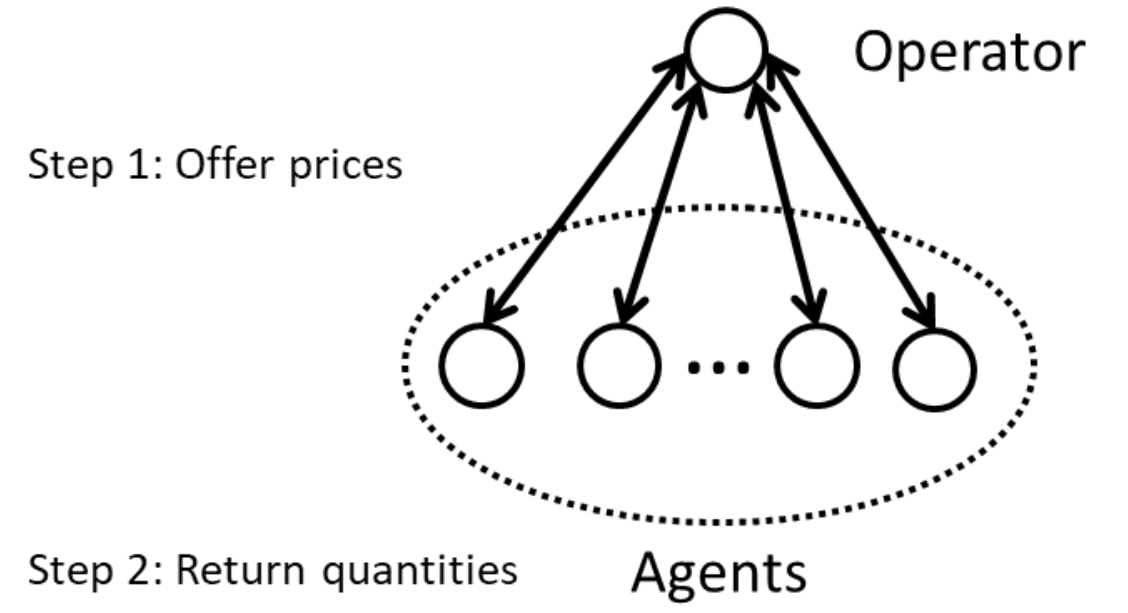}\label{fig:category2}}
\hfil
\subfigure[]
{\includegraphics[width=0.48\columnwidth]{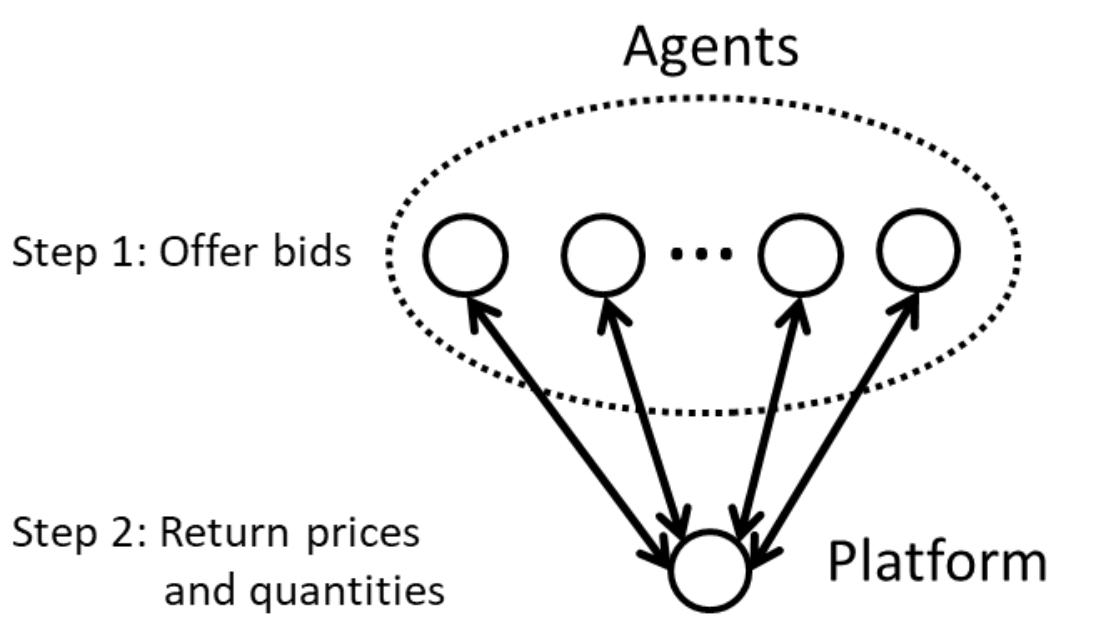}\label{fig:category3}}
\subfigure[]
{\includegraphics[width=0.38\columnwidth]{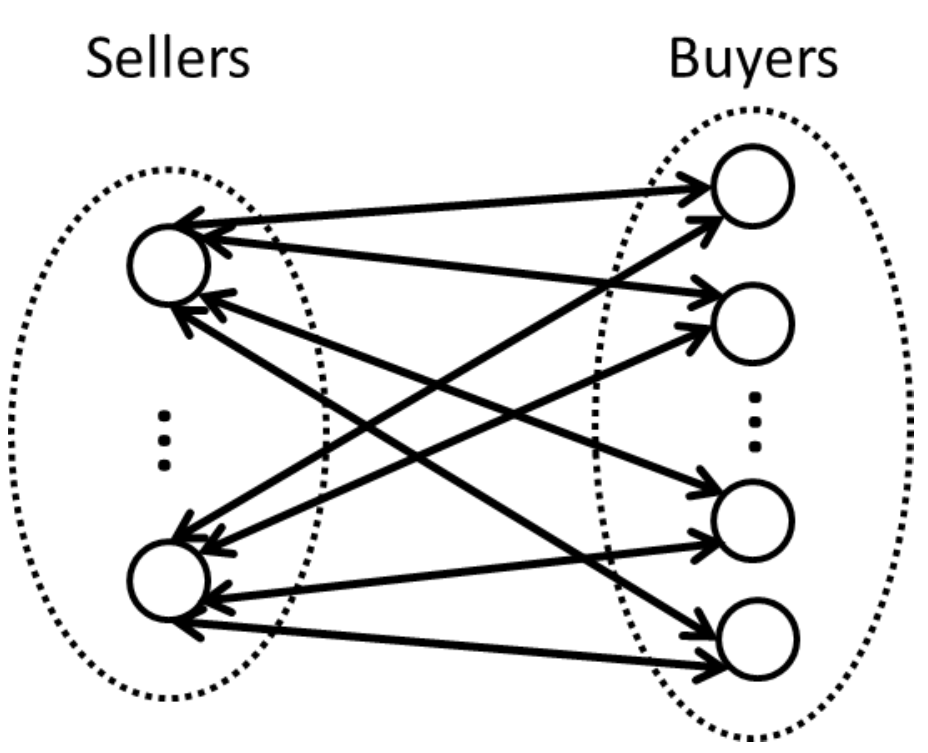}\label{fig:category4}}
\caption{Energy sharing based on (a) a Stackelberg game, (b) a generalized Nash game, and (c) a multi-leader multi-follower game.}\label{fig:category}
\vspace{-1.5em}
\end{figure}

% \begin{figure}[h]
% 	\centering
% 	\includegraphics[width=0.5\columnwidth]{Figs/category2.pdf}
% 	\caption{Stackelberg game-based energy sharing.}
% 	\label{fig:category2}
% \end{figure}

A Stackelberg game adopts the centralized leader-follower structure in Fig. \ref{fig:category2}.
%and follows a bi-level sequential decision-making procedure. 
%In the upper level, 
The operator (leader) sets prices to optimize its objective such as profit; the agents follow the prices to decide sharing quantities to maximize their net utilities.  
Specifically, consider prices $p_i^b$ and $p_i^s$ to buy and sell, respectively, with the main grid. Then the operator solves:
\begin{eqnarray}
\label{eq:category2-operator}
    \mathop{\max}_{\lambda^s,\lambda^b} 
    \begin{cases}
    \sum_{i \in \mathcal{B}} \lambda_i^b q_i + \sum_{i \in \mathcal{S}} \lambda_i^s q_i - p_i^b \sum_{i \in \mathcal{I}} q_i,~  \sum_{i \in \mathcal{I}} q_i \ge 0 \\
    \sum \nolimits_{i \in \mathcal{B}} \lambda_i^b q_i + \sum \nolimits_{i \in \mathcal{S}} \lambda_i^s q_i - p_i^s \sum_{i \in \mathcal{I}} q_i, ~ \sum_{i \in \mathcal{I}} q_i < 0 
    \end{cases}
\end{eqnarray}
and every agent $i \in \mathcal{I}$ solves:
\begin{align}
\label{eq:category2-agent}
    \mathop{\max}_{q_i}
    \begin{cases}
    v_i(q,\theta_i) - \lambda^s_i q_i & i \in \mathcal{S} \\
    v_i(q,\theta_i) - \lambda^b_i q_i & i \in \mathcal{B}
    \end{cases}
\end{align}
where $\lambda_i^s$ and $\lambda_i^b$ are the selling and buying prices of agent $i$.
%so that it receives monetary transfer $m_i= -\lambda^s_i q_i$ if $q_i\le 0 $ or $m_i = -\lambda^b_i q_i$ otherwise. 
%When $\sum \nolimits_{i \in \mathcal{I}} q_i \geq 0$, the operator needs to buy energy from the main grid to balance the market, and vice versa. 
The operator in \eqref{eq:category2-operator} maximizes its total revenue from the agents minus its net payment to the main grid, while each agent $i$ in \eqref{eq:category2-agent} maximizes its utility minus its net payment to the operator. Besides numerous heuristics, a prevalent method to compute the market equilibrium is to replace every agent’s problem \eqref{eq:category2-agent} with its optimality condition such as the Karush-Kuhn-Tucker (KKT) condition, which leads to a mixed-integer program.

Prior work along this line spans applications including microgrids with PVs \cite{liu2017energy,liu2018energy}, demand response \cite{liu2017energy2}, multi-energy systems \cite{liu2018hybrid}, multi-regional energy sharing \cite{chen2020peer}, and two-stage robust energy sharing with uncertain generation and prices \cite{cui2018two}. 
%The Stackelberg game-based energy sharing has been recognized as an effective way to accommodate renewable energy, especially in microgrids with PVs \cite{liu2017energy,liu2018energy}. 
Theoretically, \cite{cui2017distributed} proved existence and uniqueness of the Stackelberg equilibrium, and \cite{tushar2020grid} further proved stability of the equilibrium under follower coalitions.
%Reference \cite{cui2018two} proposed a two-stage robust energy sharing mechanism considering the uncertainties of renewable generation and prices. 
Moreover, efficient algorithms have been developed to reach the market equilibrium. For instance, an online algorithm was designed based on Lyapunov method to maximize the self-sufficiency of prosumers and ensure the stability of nano-grid clusters \cite{liu2017online}; a consensus-type distributed algorithm was developed in \cite{liu2018distributed}; a Q-learning algorithm was developed by embedding the prediction of rooftop PV generation \cite{xu2020data}.
%The work above made a perfect rationality assumption, which can be relaxed with subjective models of prosumers based on prospect theory \cite{chen2020peer2}. 

Besides the reliance on a central operator and the difficulty to decide optimal prices, a major limitation of the Stackelberg game is that prosumers are simply modeled as price-takers.

\textbf{2.2) Generalized Nash game}

%In the above mechanisms, the operator decides the contract or prices first, then the agents change their behaviors accordingly. In other words, the agents are price-takers. In the following two kinds of mechanisms, the agents are price-makers, i.e. they make their bids first and the prices or matching are determined afterwards.
% \begin{figure}[h]
% 	\centering
% 	\includegraphics[width=0.5\columnwidth]{Figs/category3.pdf}
% 	\caption{Generalized Nash game-based energy sharing.}
% 	\label{fig:category3}
% \end{figure}

In a generalized Nash game, the strategy set of an agent depends on the strategies of others.
It adopts the distributed structure in Fig. \ref{fig:category3}, where every agent submits a bid considering other agents' reactions, and then the platform clears the market. In particular, every agent $i\in\mathcal{I}$ maximizes  $v_i(q(b),\theta_i) -\lambda_i(b) q(b)$, i.e., its utility $v_i$ minus its net payment $\lambda_i q$ to the market, over its bid $b_i$.
Upon receiving all the bids $b:=(b_i,~\forall i \in \mathcal{I})$, the platform decides prices $\lambda=(\lambda_i,~\forall i \in \mathcal{I})$ and quantities $q=(q_i,~\forall i \in \mathcal{I})$ to minimize a social cost $P(\lambda(b),q(b))$. 
Such a generalized Nash game boils down to an equilibrium problem with equilibrium constraints (EPEC).

Reference \cite{le2020peer} modeled a distributed P2P energy market as a generalized Nash game and proved equivalence between its variational equilibria and social optima. Reference \cite{chen2020energy} proposed a generalized demand function bidding mechanism to attain a unique Nash equilibrium that approaches the social optimum with an increasing number of prosumers. 
Based on that, a practical bidding process to reach the generalized Nash equilibrium was developed in \cite{chen2020approaching}, with network power flow limits further incorporated in \cite{chen2021energy}. These mechanisms do not require complicated allocation rules or impose exogenous restriction on buyer and seller roles. 
However, they may suffer inefficiency due to the market power of some participants. 

\textbf{2.3) Multi-leader multi-follower game}

% \begin{figure}[h]
% 	\centering
% 	\includegraphics[width=0.35\columnwidth]{Figs/category4.pdf}
% 	\caption{Multi-leader multi-follower game-based energy sharing.}
% 	\label{fig:category4}
% \end{figure}

In such a mechanism, the agents are predetermined as sellers or buyers and act in a distributed market as shown in Fig. \ref{fig:category4}. The sellers first announce prices taking into account buyer reactions, followed by the buyers who decide energy sharing quantities.  
Specifically, each seller $i \in \mathcal{S}$ solves:
\begin{align}\nonumber
    \mathop{\max}_{\lambda_{ij},\forall j \in \mathcal{B}}~ & v_i(-\sum \limits_{j \in \mathcal{B}} q_{ij}(\lambda),\theta_i)+\sum \limits_{j \in \mathcal{B}} \lambda_{ij}q_{ij}(\lambda)
\end{align}
and each buyer $j \in \mathcal{B}$ solves:
\begin{align}\nonumber
     \mathop{\max}_{q_{ij},\forall i \in \mathcal{S}}~ & v_i(\sum \limits_{i \in \mathcal{S}} q_{ij},\theta_j) -\sum \limits_{i \in \mathcal{S}}\lambda_{ij}q_{ij}
\end{align}
where buyer $j \in \mathcal{B}$ purchases energy quantity $q_{ij}$ from seller $i \in \mathcal{S}$ at price $\lambda_{ij}$. A seller (buyer) aims to maximize its value plus (minus) its revenue (payment). 
Reference \cite{paudel2018peer} modeled the dynamics of buyers selecting sellers as an evolutionary game. Reference \cite{anoh2019energy} grouped physically separated prosumers into logical virtual microgrids and modeled their energy sharing as a multi-leader multi-follower game. Reference \cite{lee2015distributed} designed an alternative procedure that buyers bid prices and sellers decide quantities.
A major limitation of these mechanisms is the inflexibility of market roles of prosumers.

\textbf{2.4) Bilateral Nash game}

Such a mechanism has a similar structure to that in Fig. \ref{fig:category4}, but works in a quite different way. 
%Under this kind of energy sharing mechanism, trading offers are posted and handshakes are made. 
%First, each agent is registered as a seller or a buyer.
The pre-assigned sellers and buyers post their offers, and the sharing quantities are determined via matching or auctions. Then the bilateral prices or contracts are settled \cite{liu2019peer}.
A key advantage of this mechanism is its compatibility with asynchronous actions, and a key challenge is to find matching results in a scalable way. 

A bilateral contract network in forward and real-time markets was proposed in \cite{morstyn2019bilateral}. A P2P energy transaction matching algorithm was developed in \cite{oh2020peer} considering uncertainty and fairness of profit allocation. A negotiation algorithm settled an agreement among prosumers without prior knowledge about the preference of each other \cite{khorasany2020new}. Reinforcement learning was used by an energy broker to match sellers and buyers \cite{chen2018indirect}. Energy user behavior can be modeled using reinforcement learning with bounded regret and embedded into the mixed-integer linear program for seller-buyer matching \cite{agate2020enabling}. Double auctions were also held for bilateral matching \cite{he2020community,wang2014game}. 

\subsection{Optimization models}

Optimization-based energy sharing mechanisms aim to reach social optimum. However, solving the social optimization in a centralized manner would compromise the privacy of participants.
To address that, many studies interpreted dual variables of the social optimization as energy prices and carried out distributed iterative processes for energy sharing, which converge to socially optimal market equilibria. 

The bidirectional energy exchange between EVs and roads was conducted with a privacy-preserving consensus algorithm \cite{nguyen2020electric}. Reference \cite{nguyen2020cooperative} further considered strategic choice of objective function parameters by the agents. Other approaches include the gradient algorithm \cite{le2018enabling} and the alternating direction method of multipliers \cite{yang2020transactive,paudel2019pricing}.
Besides, learning algorithms have gained popularity for energy sharing. A modified deep Q-network was applied to microgrids that share energy to maximize their total utility \cite{chen2019realistic}. 
Multi-agent deep reinforcement learning aided energy sharing to realize a zero-energy community \cite{prasad2019multi}. 
%This kind of mechanisms can achieve social optimal outcome. 
A disadvantage of optimization-based designs lies in the difficulty to reveal the underlying economic intuition and incentivize the agents to participate.

\section{Future Research}
%Energy sharing aims to enhance resource utilization efficiency via exchange. In this paper, the application scenarios, some emblematical business models, and the mechanism design methods are summarized. 
This section discusses three directions to explore about energy sharing, a promising concept that is still in its infancy.
 %the participants (including the operator and the agents), the energy/information/value flows, and the environment.

\subsection{Participants}
1) \emph{Operators.} Most existing work assumes only one operator in an energy sharing system. Market designs and pricing strategies with more than one competing operators require further investigation. Another critical issue is for an operator to attract customers and maintain their loyalty.

2) \emph{Prosumers.}
%Before participating in energy sharing, the agents need to choose which sharing group to join. 
It is valuable to group the prosumers for better energy sharing outcomes. Despite the various clustering algorithms \cite{bedingfield2018multi}, most of them simply bundles prosumers with similar utility functions or load patterns which may not fully realize the potential of sharing. Especially, if the prosumers in a group are identical, sharing would not create much benefit. 
Besides clustering, the matching between prosumers \cite{morstyn2019bilateral} as well as prosumer's bounded rationality that can be analyzed with the prospect theory \cite{saad2016toward, el2017managing, chen2020peer2} are also crucial.

\subsection{Energy, information, and value flows}
1) \emph{Energy.} The work so far has mainly concentrated on sharing electricity. As co-generation technologies are connecting multiple energy systems tighter than before, the sharing and conversion between multiple energy forms and the development of an integrated energy market will become inevitable.

2) \emph{Information.} Energy sharing has mostly been studied under a symmetric information structure where the participants have full and equal knowledge about each other. However, asymmetric information is more common in reality \cite{nguyen2020cooperative}, which may cause market failure if not handled properly. 
In particular, it deserves efforts to design mechanisms that can motivate participants to truthfully report their relevant information. Furthermore, when user data such as load profiles are needed for market operation, the information privacy and the value of information \cite{wang2021data} are issues of great merit.

3) \emph{Value.} Energy sharing markets today are mostly settled with financial support. It is worth the attempt to apply other paradigms, such as the credit score system for carbon and emission trading \cite{show2008carbon}, to energy sharing.

\subsection{Environment}
The uncertainty of large-scale renewable generation is a key challenge for energy systems. Classical methods to overcome this challenge include stochastic, robust, and distributionally robust optimizations. Those methods often assume uncertainty scenarios or sets that are modeled exogenously. 
However, such uncertainty can actually depend on operating decisions. For example, the changing angle of a PV panel can influence its power output. This kind of decision-dependent uncertainty has not received adequate attention, partly due to its computational intractability that needs to be addressed in the future \cite{nohadani2018optimization}.

\section{Conclusion}
This paper serves as a comprehensive survey of energy sharing from its basic structures, applications, business models, and market mechanisms. Its three typical structures are centralized, distributed, and decentralized. Applications of energy sharing can accommodate volatile renewable sources such as solar, wind, and hydrogen; enhance the operating efficiency of smart buildings, microgrids, and integrated energy systems; and relieve the reliance on energy storage backups. 
We reviewed business models of energy sharing from perspectives of resource sharing modes and flexibility levels.
Market mechanisms were categorized into game-theoretic and optimization models. Finally, future research directions were discussed in terms of operators and prosumers; energy, information, and value flows; and uncertainty of the environment.

	\ifCLASSOPTIONcaptionsoff
	\newpage
	\fi
	
	\bibliographystyle{IEEEtran}
	\bibliography{IEEEabrv,mybib}

\end{document}